\documentclass[12pt]{article}

\usepackage{amssymb}  
\usepackage{amscd} 
\usepackage{wrapfig}
\usepackage{graphicx}
\usepackage{cite}

\newcommand {\newsection}{\setcounter{equation}{0}\section}

\def \cc {{\mathbb C}} 
 
\def \de {\partial} 
 
\def \nn {{\cal N}}
\def \PP {{\cal P}}

\begin{document}
\rightline{Bicocca-FT-03-10}
\rightline{IHES/P/03/27}
\rightline{RR 017.04.03}
\vskip .1in \hfill hep-th/0304251

\hfill

\vspace{20pt}

\begin{center}
{\Large \textbf{On the Geometry of Matrix Models for N=1*}}
\end{center}

\vspace{6pt}

\begin{center}
\textsl{M. Petrini $^{a}$, A. Tomasiello $^b$, A. Zaffaroni $^{c}$} 
\vspace{20pt}

\textit{$^a$ Centre de Physique Th{\'e}orique, Ecole Polytechnique, 48 Route de Saclay; F-91128 Palaiseau Cedex, France.}

\vspace{10pt} 

\textit{$^b$ I.H.E.S.,
Le Bois-Marie, Bures-sur-Yvette, 91440, France.}

\vspace{10pt} 

\textit{$^c$ Universit{\`a} di Milano-Bicocca and INFN\\ Piazza della 
Scienza 3, I-20126 Milano, Italy.}

\vspace{10pt} 

\end{center}

\vspace{12pt}

\begin{center}
\textbf{Abstract}
\end{center}

\vspace{4pt} {\small \noindent 
We investigate the geometry of the matrix model associated with 
an $\nn=1$ super Yang-Mills theory with three adjoint fields,
which is a massive deformation of $\nn=4$.
We study in particular the Riemann surface underlying solutions
with arbitrary number of cuts.
We show that
an interesting geometrical structure emerges where
the Riemann surface is related  on-shell to the Donagi-Witten 
spectral curve. We explicitly identify the quantum field theory 
resolvents in terms of geometrical data on the surface.
}
%\vfill\eject
%\noindent
\vfill
\vskip 5.mm
 \hrule width 5.cm
\vskip 2.mm
{\small
\noindent e-mail: Michela.Petrini@cpht.polytechnique.fr, 
Alessandro.Tomasiello@cpht.polytechnique.fr, 
alberto.zaffaroni@mib.infn.it}

\newpage

\newsection{Introduction}

Recently it has been proposed to use matrix models to extract information
about holomorphic quantities of a certain class of $\nn=1$ gauge
theories. More precisely,
%Some recent works have proposed a relation between 
%$\nn=1$ gauge theories and matrix models. 
the exact superpotential
and the condensates of chiral operators in the gauge theory can be
computed from the free energy of an associated matrix model \cite{dv3}.
This proposal has been extensively tested in the case 
of a pure $U(N)$ theory with an adjoint field with potential
$W(\Phi)$ (and possibly matter in the fundamental
representation), showing that the matrix
model results and the quantum field theory ones agree. 
%The case of a pure $U(N)$ theory with an adjoint field with potential
%$W(\Phi)$ (and possibly matter in the fundamental
%representation) is relatively well understood. This case 
%has been extensively studied, showing that the matrix
%model results and the quantum field theory ones agree. 
This model has 
an interesting
geometrical structure in terms of a Riemann surface \cite{int,cv,cdsw,csw3}.
%emerges by studying the Ward identity for the model. 
%From the analysis of the pure $U(N)$ theory
In particular, the resolvents in the matrix
model and in the field theory seem to be related to
geometrical quantities on this surface. This geometrical
structure is deeply related to the Seiberg-Witten curve of the 
$\nn=2$ theory of which the $\nn=1$ theory is a deformation.
In this paper we will focus on the geometry of the $\nn=1$ theories
that are obtained as deformations of the $\nn=4$ SYM. More precisely,
we consider the model with three adjoint fields, a mass term for two
of them and a generic potential for the third one. The associated
matrix model was solved in the case of a single cut in \cite{dv3,dhks},
showing remarkable agreement with the quantum field theory results.
In this paper, we are interested in the geometry of the 
matrix model for arbitrary
number of cuts. Since the model can be considered as an $\nn=1$ 
deformation of an $\nn=2$ theory with a massive hypermultiplet,
the corresponding geometry is expected to be related to
the Donagi-Witten curve \cite{dw}.   The emergence of an elliptic curve 
in the case of a single cut \cite{dv3} is a first example of this
connection. We will investigate the relation between the matrix model and 
the Donagi-Witten curve, 
and the geometrical structure that emerges in this way.

To be more concrete, 
we will study an $\nn=1$ supersymmetric gauge theory with
gauge group $U(N)$ and three chiral fields $\Phi$, $X$, $Y$ 
in the adjoint representation. The theory is characterized
by the superpotential
\begin{equation}
  \label{eq:supot}
  {\cal W}(\Phi, X, Y)= i\Phi [X, Y] + m X\,Y + W(\Phi)\ ,
\end{equation}
where $W=\sum_{k=1}^{n+1}g_k{\rm Tr}\Phi^k$ 
is a polynomial in $\Phi$ of degree $n+1$. 
This model can be seen as a deformation of the $\nn=2$
theory with a massive adjoint hypermultiplet considered by
Donagi and Witten \cite{dw}.
We are mainly interested in a
classical vacuum  where $X$ and $Y$ vanish and $W'(\Phi)=0$,
even though  our results can be applied also to other vacua. 
The $N$ classical 
eigenvalues of
$\Phi$ can be distributed over the $n$ points $\phi_i$ where $W'(x)=
\prod_{i=1}^n (x-\phi_i)=0$. 
The gauge group is correspondingly broken to $\prod_{i=1}^{n} U(N_i)$, 
where $N_i$ is the number of
eigenvalues corresponding to the $\phi_i$ critical point of $W$. 
At low energy, we can write an effective superpotential 
\begin{equation}
W_{\rm eff}(S_i,\phi_i,m,\tau)
\end{equation}
for the condensate superfields 
$S_i={\rm Tr} (W_{\alpha}W^{\alpha})_{(i)}$.

In quantum field theory, the vacuum expectation value of
$W_{\rm eff}$ can be computed (in principle) from the knowledge of
the underlying $\nn=2$ Seiberg-Witten curve. The addition of the
superpotential $W(\Phi)$ indeed selects particular points in the
moduli space of the $\nn=2$ theory, which can be found by
explicitly solving the equations of motion. Once these points
are found, $W_{\rm eff}$ can be determined via
\begin{equation} 
W_{\rm eff}=\langle W(\Phi)\rangle_{\nn=2}
=\sum_{k=1}^{n+1}g_k{\rm Tr}\langle\Phi^k\rangle_{\nn=2} \ .
\end{equation}
In our case the $\nn=2$ theory is described by the Donagi-Witten curve 
for $U(N)$, which is an $N$-sheeted covering of a torus.
The points selected by the equations of motion are typically
points of partial or maximal degeneracy of the curve. In the 
generic vacuum $\prod_{i=1}^{n} U(N_i)$ we indeed expect that
$N-n$ monopoles are massless. 

In the corresponding matrix model, the vacuum
$\prod_{i=1}^{n} U(N_i)$ is associated with an $n$-cut
solution of the equations of motion. We will show that, 
as in \cite{dv3}, the matrix model can be associated with a 
Riemann surface of genus equal to the number of cuts.
$W_{\rm eff}$ is then computed from geometrical data on this surface.
It is then interesting to investigate the relation of this
geometrical structure with the Donagi-Witten curve.
The single cut case was solved in \cite{dv3,dhks}. 
It corresponds to a field theory massive vacuum where 
the $\nn=2$ spectral curve has maximal degeneracy
and itself becomes a torus with modular parameter $\tau/N$. 
The emergence of such torus on the matrix model side was explicitly 
determined in  \cite{dv3,dhks}. 
The connection with the $\nn=2$ curve becomes more explicit when
one considers the opposite case of maximal number of cuts.
As in \cite{cv}, we can introduce a degree $N+1$
superpotential with 
$W'(x)=
\epsilon \prod_{i=1}^N (x-\phi_i)=0$. In the vacuum where
all the $N$ eigenvalues of $\Phi$ are distinct the gauge group
is broken to the maximal abelian subgroup $U(1)^N$.  
In the limit where the superpotential is turned off ($\epsilon\rightarrow 0$), 
we should recover the dynamic of
the $\nn=2$ theory. We will show that the Riemann surface of
the associated matrix model 
with $N$ cuts becomes in this limit the Donagi-Witten curve.

We will also analyze the geometry of the matrix model 
for an arbitrary number of cuts.
We will show that, upon minimization,
the Riemann surface always becomes a covering of a torus
and we will discuss the relation of this surface
with the Donagi-Witten construction. 
We will also identify the field theory
resolvents with geometrical quantities on the
Riemann surface. All the results have 
a direct analogue in the case of a pure $U(N)$ theory 
\cite{dv3,cv,cdsw,csw3}. 
%The only result that we will not be able to write
%explicitly is a complete set of loop equations, since these are
%considerably more complicated than in the case of pure gauge.
%We will nevertheless derive some suggestive relations. 
We will mainly consider 
the on-shell theory, where a minimization with respect to the moduli
has been performed. 
The organization of the paper is as follows. In Section 2, we discuss
the Riemann surface associated with the off-shell theory. In Section 3,
we will discuss the conditions following from the minimization and
the relation of the on-shell theory with the Donagi-Witten curve. In
Section 4, we discuss the identification of the field theory
resolvents with geometrical quantities on the Riemann surface.
Finally, in the Appendix we briefly discuss the loop
equations and some identities for the off-shell theory.

\newsection{The Riemann surface} 
Following \cite{dv3}, we can compute $W_{\rm eff}$ from
 the large $\hat N$ expansion of a matrix model:
\begin{equation}
\int {\cal D} \hat\Phi {\cal D} \hat X {\cal D} \hat Y e^{ \frac{1}{g_s} 
{\rm Tr} \{
i \hat\Phi [\hat X, \hat Y] + m \hat X \, \hat Y + W(\hat\Phi)\}},
\label{path}
\end{equation}
where  $\hat \Phi, \hat X, \hat Y$ are $\hat{N} \times \hat{N}$ hermitian matrices. The model can be solved integrating out the matrices 
$\hat X, \hat Y$ and diagonalizing the  remaining matrix 
$\hat \Phi$. 
The saddle point 
equation of motion reads
\begin{equation}
  \label{eq:mmeom}
W'(\lambda_I) = g_s \sum_{J\neq I}\left[ 
2\frac1{\lambda_I-\lambda_J}
-\frac1{\lambda_I-\lambda_J+ im}
-\frac1{\lambda_I-\lambda_J- im}\right]\ ,
\end{equation}
where $\lambda_I$ are the eigenvalues of $\hat\Phi$.
As usual, in the large $\hat{N}$ limit, the eigenvalues will be spread 
over $n$ cuts around each solution of $W'(\hat\Phi)$. We will
denote the fraction of eigenvalues for each cut as $\hat N_i$.
According to the Dijkgraaf--Vafa prescription the
filling fractions $S_i=g_s\hat N_i$ are identified with the field theory
condensates and the effective superpotential 
corresponding to  the $\prod_{i=1}^{n} U(N_i)$ vacuum is \cite{dv3} 
\begin{equation}
  \label{eq:Weff1}
  W_{\rm eff}= 
\sum_i \left( N_i \frac{\de {\cal F}}{\de S_i} - 2\pi i\tau S_i \right),
\end{equation}
where ${\cal F}(S_i)$ is the matrix model free energy. 
Unless explicitly stated, we will
consider the case with the maximal allowed number of cuts.

Information about the model is encoded in the resolvent
$\omega(x)\equiv \frac{1}{\hat{N}}{\rm Tr} \frac{1}{x-\hat\Phi}$. 
As usual, $\omega(x)$ has cuts corresponding to the distribution of the 
matrix model eigenvalues.
For this particular model, 
it is useful to  define \cite{nek,dv3,dhks} the function 
\begin{equation}
  \label{eq:G}
  G(x) = U(x) + 
i S \left[ \omega\Big(x+\frac i2 \,m\Big) -
\omega\Big(x-\frac i2 \,m\Big)\right]\ ,
\end{equation}
where $S\equiv g_s \hat{N}$ is the 't Hooft coupling and 
$U(x)$ is defined by
the property
\begin{equation}
  \label{eq:U}
  U\Big(x+\frac i2\,m\Big)-U\Big(x-\frac i2\,m\Big) = W'(x)\ .
\end{equation}
Then as a consequence of (\ref{eq:mmeom})  \cite{nek,dv3,dhks}
\begin{equation}
  \label{eq:rs}
G\Big(x+\frac i2 \,m \pm i\epsilon\Big)=
G\Big(x-\frac i2 \,m \mp i\epsilon\Big)\qquad {\rm for}
\ \, x \ \, {\rm on\  a\ cut.}
\end{equation}
%This is not the only consequence of the matrix model equations of
%motion (\ref{eq:mmeom}). One can also extract loop equations
%that we will consider later in section \ref{sec:res}. For the time
%being we stick to (\ref{eq:rs}). 
%This equation is important because it
%already displays the $\nn=1$ curve.

%\begin{wrapfigure}[8]{R}{7cm}
%\hspace{.5cm}
%\begin{picture}(80,50)
%\put(180,50){\small\fbox{$x$}}
%\includegraphics[width=7cm]{mm1.eps}
%\end{picture}
%\caption{\small Cuts in the $x$ plane.}
%\label{cuts}
%\end{wrapfigure}
%
%\begin{wrapfigure}[8]{R}{7cm}
%\hspace{.5cm}
%
%\begin{figure}[h]
%\centerline{{\small\fbox{$x$}}
%\includegraphics[width=13cm,height=4.5cm]{mm.eps}}
%\caption{\small Cuts in the $x$ plane for the function $G$.
%Lower cuts are identified with upper cuts as shown using dashed lines.
%$A_i$ and $B_i$ form a basis of cycles for $\Sigma$.}
%\label{cuts}
%\end{figure}

\begin{figure}[h]
\hspace{.5cm}
\begin{picture}(80,140)(-10,0)
\put(0,0){{\small\fbox{$x$}}}
\put(15,0){\includegraphics[width=13cm,height=4.5cm]{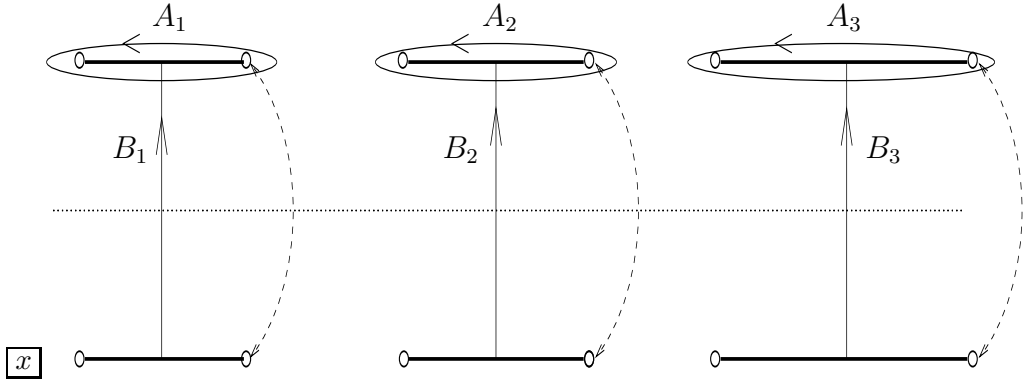}}
\put(40,80){$B_1$}\put(55,130){$A_1$}
\put(165,80){$B_2$}\put(180,130){$A_2$}
\put(325,80){$B_3$}\put(310,130){$A_3$}
\end{picture}
\caption{\small Cuts in the $x$ plane for the function $G$.
Lower cuts are identified with upper cuts as shown using dashed lines.
$A_i$ and $B_i$ form a basis of cycles for $\Sigma$.}
\label{cuts}
\end{figure}

Equation (\ref{eq:rs}) means that $G$ is well-defined
on a surface obtained from the $x$ plane after $n$ identifications, as
shown in figure \ref{cuts}. If one adds the point $x=\infty$, 
this space is topologically 
a Riemann surface $\Sigma$ 
of genus $n$. Notice that this way of describing a Riemann surface
is somewhat reminiscent of the light cone parameterization of moduli
space of punctured Riemann surfaces \cite{gw}.

The function $G$ and the Riemann surface $\Sigma$ allow to write
the effective superpotential~(\ref{eq:Weff1}) in a more convenient form.
First of all, integrating equation~(\ref{eq:G}) around the cuts
we obtain the relation
\begin{equation}
S_i= \frac1{2\pi}\int_{A_i} G\, dx\ ,
\label{Si}
\end{equation}
where $A_i$ are the cycles encircling the cuts in the $x$ plane.
Moreover, as shown in \cite{dv3}, $\partial {\cal F}/ \partial S_i$
can be written in terms of integrals over the cycles $B_i$ going
from a lower cut to an upper one. The superpotential then reads
\begin{equation}
  \label{eq:Weff}
  W_{\rm eff}= 
i \sum_i \left( N_i \int_{B_i} G\, dx - 
    \tau \int_{A_i} G\, dx \right).
\end{equation}

We are interested in the geometrical structure of the problem. 
To this purpose, we can re-formulate the matrix model data 
in the following way. 
The coordinate $x$ of the matrix model plane is not a well-defined 
function on the Riemann surface $\Sigma$. Its differential $dx$, though, 
is well-defined. It has a double pole in the point $x=\infty$ and 
is regular 
otherwise; its $A$ periods are zero, and its $B$ periods are $i\,m$.
Moreover we can see from eq. (\ref{eq:G})
that the function $G$ has a single pole of order $n+1$ 
at the point $x=\infty$. By Riemann-Roch this property
singles out $G$ up to an additive constant. 
Thus our geometrical data are 
\begin{itemize}
\item a Riemann surface $\Sigma$ of genus $n$ 
\item a differential $dx$ with a double pole and periods 
  \begin{equation}
    \label{eq:dx}
    \int_{A_i} dx =0\ ,\qquad\int_{B_i} dx = i\,m\ .
  \end{equation}
\end{itemize}

Let us count the number of moduli of our data. 
We have $3n-3+1$ moduli from the moduli
space of Riemann surfaces of genus $n$ with one puncture. Again 
from Riemann-Roch one gets $n+1$ for the number of 
differentials with a double pole; integrating this to obtain $x$ 
involves an extra additive constant. Taking away $2n$ from (\ref{eq:dx}) 
one gets finally 
\[
(3n-2)+ (n+1)+1-(2n)= 2n\ .
\]

From the matrix model point of view, 
these $2n$ moduli are easily interpreted as the classical vacua
(points $\phi_i$ in which $W'(\phi_i)=0$),  
and the filling fractions $S_i$. 
%Physically, the
%latter are gaugino condensates relative to the $N$ factors $SU(N_i)$. 
%As for 
%the $x_i$, let us consider as in \cite{cv} an $\nn=2$ 
%limit in which they are kept 
%constant, while sending to zero an overall constant multiplying the 
%superpotential ($W'(x)=\epsilon w'(x)$, $\epsilon\to 0$). Then the $x_i$ get
%interpreted as moduli of the resulting $\nn=2$ theory, that is the 
%Donagi-Witten as we stressed at the beginning. 
%Let us count the number of moduli of our data. 
%Before we consider that in more detail in next section, 
Alternatively, we can also interpret the $2n$ moduli as
%some more feature of the data $(\Sigma, dx)$ considered above. 
%First of all,
the zeros of $dx$. A meromorphic 
differential with a double pole must indeed have $2n$ zeros: 
in the $x$ plane 
they are the end-points of the cuts, denoted by small circles 
in figure \ref{cuts}. We see
then that $dx$ has precisely the role played by $dw$ in the light-cone 
parameterization of the moduli space of Riemann surfaces with punctures in
\cite{gw}\footnote{The condition of having specified periods has an analogue 
in that case, where the differential was required to have purely
imaginary periods.  A differential with a double pole and purely imaginary
periods is uniquely determined on any Riemann surface
by a procedure similar to that in \cite{gw}; 
then imposing the actual value of these
periods as in (\ref{eq:dx}) constrains to a sub-variety of the moduli space.
The only real difference with \cite{gw} is that $dw$ has two single
poles while our $dx$ has one double pole.}. 

It is interesting to compare our data
with the case of a pure $U(N)$ gauge theory. In that case,
the Riemann surface has genus $n$ and can be explicitly written as
\begin{equation}
\label{riemann1}
y^2=W'(x)^2+f_{n-1}(x) \ ,
\end{equation}
where $f_{n-1}$ is a polynomial of degree $n-1$, whose coefficients
are related to the moduli $S_i$. The role of the meromorphic function $G$ 
is played by $y$: all the relevant formulae are obtained with  
the substitution
$Gdx \rightarrow ydx$.
An important difference with respect to our case,
is that the matrix model plane for pure $U(N)$ corresponds to just
one sheet of the surface~(\ref{riemann1}). On this sheet we have
the relation
$y=W'(x)-2\omega(x)$, instead of equation~(\ref{eq:G}).
All quantities are then continued to the second sheet. In our case,
the matrix model plane is topologically identified with the entire
Riemann surface.

\section{Minimizing the superpotential}

The results obtained so far from the matrix model apply to
an off-shell theory. In order to obtain
the  vacuum value of the superpotential we further have to 
minimized equation~(\ref{eq:Weff}) with respect to $S_i$.
We refer to the theory after minimization as the on-shell theory.
%We repeat here the idea of this limit \cite{cv}. Choosing a superpotential
%$W(\Phi)=\epsilon \Pi_i (x -x_i)$ and sending $\epsilon\to 0$ one can find 
%oneself in an arbitrary point of $\nn=2$ moduli space. {\bf Da spiegare 
%meglio.} Here we are also {\bf supposing (?)}  that the $x_i$ 
%are not modified quantum mechanically, 
%that in the pure $\nn=2$ gauge case was 
%established \cite{cv} thanks to explicit result that we lack here.
%So we have to minimize the superpotential $W_{\rm eff}$ with 
%respect to $S_i$. 

%We will see that after minimization the Riemann surface $\Sigma$
%is related to the Donagi-Witten curve. 

Define the differentials
\begin{equation}
\omega_i=\frac{1}{2\pi}\frac{\de}{\de S_i} G\,dx\ .
\label{differentials1}
\end{equation}
The $\omega_i$ have no poles and 
are therefore holomorphic differentials.
%Indeed, when we differentiate equation~(\ref{eq:G}), $\omega_i$ 
%get contributions
%only from the resolvent $\omega(x)$, which behaves at infinity
%as $\omega(x)\sim 1/x$; the potential simple pole is canceled 
%in the difference of the two resolvents
Indeed, when we differentiate equation~(\ref{eq:G}), the $\omega_i$ 
get contributions
only from the difference of the resolvents; 
the simple pole in the resolvent, which behaves at infinity
as $\omega(x)\sim 1/x$, cancels 
in the difference \footnote{Notice the difference
with the case of pure gauge \cite{cv,dv3} where one of the
derivatives of $ydx$ with respect to the $S_i$ is a meromorphic
differential with a simple pole. This difference is consistent with
the fact that $\Sigma$ has genus$N$, while the hyperelliptic curve
considered in \cite{cv,dv3} has only genus $N-1$.}. 
From equation~(\ref{Si}) it also
follows that the $\omega_i$ form a basis of canonically normalized
holomorphic differentials: $\int_{A_i}\omega_j=\delta_{ij}$. Minimizing
equation~(\ref{eq:Weff}) with respect to $S_i$ we obtain
\begin{equation}
\sum_i N_i \int_{B_i} \omega_j =\int_{B_j} \sum_i N_i \,\omega_i=\tau \ ,
\label{mimim}
\end{equation}
where we used the symmetry of the period matrix $\int_{B_j}\omega_i$.

%The consequences of the minimization can be expressed in terms of
%the holomorphic differential $\Omega=\sum_i N_i \omega_i$.
The effects of the minimization translate into the behavior of
the holomorphic differential $\Omega=\sum_i N_i \omega_i$.
Its $A$-periods are $\int_{A_i}\Omega=N_i$. The minimization tells us
that, on-shell, the $B$-periods of $\Omega$ are 
all equal, $\int_{B_i}\Omega=\tau$. 
%It is actually convenient to consider as in \cite{cv} variables $S, S_{i,i+1}$
%such that $\de/\de S=\sum_i N_i \,\de/\de S_i$ and $\de/\de S_{i,i+1}$=
%$\de/\de S_i -\de/\de S_{i+1}$. In these terms the result of the minimization
%is
%\begin{equation}
%  \label{eq:min}
%  \begin{array}{c}\vspace{.2cm}
%  \sum_j N_j\int_{A_j} \frac{\de G}{\de S}dx= 1\ ,\quad
%  \sum_j N_j\int_{A_j} \frac{\de G}{\de S_{i,i+1}}dx= 0\ ; \\ 
%   \sum_j N_j\int_{B_j} \frac{\de G}{\de S}dx= \tau\ ,\quad
%  \sum_j N_j\int_{B_j} \frac{\de G}{\de S_{i,i+1}}dx= 0\ .
%  \end{array}
%\end{equation}
%From the definition of $G$ (\ref{eq:G}), 
%one sees that 
%\begin{equation} 
%\omega_0=\de G /\de S,\qquad  \omega_i=\de G/\de S_{i,i+1}, i=1,..,n-1
%\label{holo}
%\end{equation}
%are a basis of holomorphic differentials on the surface.
%(\ref{eq:min}) therefore tells us that one line of the period matrix is
%$(\tau, 0,\ldots,0)$. {\bf qualche parola in piu'}
To see what this means, consider the map
\begin{equation}
  \label{eq:z}
P\quad \mapsto \quad z(P)\equiv\int_{P_0}^P \Omega \,
%\int_{P_0}^P \frac{\de G}{\de S_{i+1,i}}dx  
\end{equation}
from the Riemann surface to $\cc$. 
%The map (\ref{eq:z})
%can be considered as an ``incomplete Jacobi map'', which, for
%a generic Riemann surface, makes no sense at all. However, in our
%case, since all the $B$-period are equal and all the $A$-period
%are integers, we see that (\ref{eq:z}) is well defined provided
%we make the identifications
The map (\ref{eq:z}) could be considered an ``incomplete Jacobi map'', in the
sense that we only consider the integral of one of the 
holomorphic differentials.
% another peculiarity is that the
%image is $\cc$ and not a compact manifold. 
For a generic Riemann surface, one would not be able to do better and identify 
points to make the image compact. However, in our
case, since all the $B$-periods are equal and all the $A$-periods
are integers, (\ref{eq:z}) is a well--defined map to a torus 
defined by the identifications
\begin{equation}
z\sim z+\tau\, \ , \qquad\qquad z\sim z+\tilde N\ , 
\label{torus}
\end{equation}
where $\tilde N$ is the highest common factor of the $N_i$.
%Since the
%period matrix is symmetric, the holomorphic
%differential $\de G/\de S$ has periods only on the cycles 
%$A\equiv\sum_i N_i A_i$ and $B\equiv\sum_i N_i B_i$. 
%If we start from one point on $\Sigma$ and follow $z$ until we are back, we
%will always find the same value if the cycle we followed is anything different
%from $A$ and $B$. If we go around these two latter cycle, $z$ does change by
%1 and $\tau$ respectively. So the map (\ref{eq:z}) is well defined if we
%identify $z\sim z+1 \sim z+\tau$, which means that the image is actually
%$T^2$. We also note that 
%(\ref{eq:z}) is then an ``incomplete Jacobi map'', and the $T^2$ we just
%described is naturally a sub-manifold of the Jacobian of $\Sigma$, a complex
%torus of dimension $N$ in which $\Sigma$ can be embedded using the 
%usual Jacobi map.
Thus, we have shown that using the equations of motion for $S_i$, 
$\Sigma$ becomes a covering of a torus of modular parameter 
$\tau/\tilde N$. 
%The cycles of this torus are $A$ and 
%$B$, and the differential $\de G/\de S \,dx= dz$. 

\subsection{The $\nn=2$ case}
To understand our result better, it is convenient to consider
first the case $n=N$, maximal number of cuts $N$ and $N_i=1$.
The gauge group is completely broken to the maximal abelian
subgroup $U(1)^N$. This is the 
situation where we expect to recover information about the
underlying $\nn=2$ theory. Following \cite{cv},
we take a potential $W(\Phi)$ of degree $N+1$ 
with $W'(x)=\epsilon \, \Pi_i (x -\phi_i)$. Since all $N_i=1$,
the eigenvalues of $\Phi$ are all distinct and 
classically coincide with the $N$ number $\phi_i$. 
The $\nn=1$ theory we are considering differs from an
$\nn=2$ theory only for the presence of
the potential $W(\Phi)$. 
If we turn off the deformation by sending $\epsilon\to 0$, 
we recover the $\nn=2$ theory in the point 
of the moduli space specified by the VEVs $\phi_i$ \cite{cv}.
%To discuss this case, we consider $n=N$ cuts
%and all $N_i=1$. In particular $\tilde N=1$.
%We will now show that, after minimization, 
%the matrix model Riemann surface $\Sigma$ becomes the Donagi-Witten
%curve describing the $\nn=2$ theory.
%
%In order to do so, we need some basic information about the Donagi-Witten
%curve \cite{dw} associated with a $U(N)$ gauge group. 

The $\nn=2$ theory we obtain in this way has gauge group $U(N)$,
a massive adjoint hypermultiplet \cite{dw}
and coupling constant $\tau$. The corresponding Seiberg-Witten curve 
%for this $\nn=2$ theory 
was determined in \cite{dw}. 
The curve can be written as an $N$-sheeted covering 
of a base torus (of modular parameter $\tau$) expressed by the equation 
\begin{equation}
F_N(v,z)={\rm det} (v-\Phi(z))=0\ ,
\label{dwcurve}
\end{equation}
where $\Phi$ is a section of a $U(N)$-Higgs-bundle on the torus.
The theory is a massive deformation of $\nn=4$ SYM and the 
presence of the torus reflects the S-duality of the original $\nn=4$
theory. Equation~(\ref{dwcurve}) gives the Donagi-Witten curve as a
degree $N$ polynomial in $v$ with coefficients that are elliptic 
functions on the base torus. 
The curve~(\ref{dwcurve}) is uniquely characterized by the existence of
a meromorphic function, $v$, with $N$ single poles 
at the $N$ counter-images $p_i$ 
of the point $z=0$ on the base torus. In one of the points, $p_0$, 
the residue has to be $-(N-1)m$, while the other $N-1$ points 
have residue $m$. The existence of a meromorphic function with these
properties uniquely determines the curve~(\ref{dwcurve}) \cite{dw,dph}.

We will now show that, after minimization, 
the matrix model Riemann surface $\Sigma$ for $n=N$ and $N_i=1$
becomes the Donagi-Witten 
curve describing the $\nn=2$ theory.
We have already seen that
the surface $\Sigma$ becomes a covering of a torus
of modular parameter $\tau$ ($\tilde N=1$ in this case). 
We will shortly see that the order $\tilde n$ of 
this covering is actually $N$. We can 
construct, on shell, the meromorphic function $v$ 
that characterizes a Donagi-Witten curve.
It is easier to find first the differential $dv$, which 
should have poles of order 2 with coefficient equal to minus the 
residue of $v$. 
We have at our disposal at least two differentials 
with double poles in some or all the points $p_i$. 
One is the pull-back of the differential $\PP(z) dz$ on the base torus. 
%where we now call $\Omega=dz$.
Once pulled back from $T^2$ to
$\Sigma$, this indeed provides double poles with coefficient one on 
every point $p_i$ over $z=0$.
Another one is $dx$ which has a double pole at $x=\infty$. 
It is natural to suppose that $x=\infty$ corresponds to $p_0$\footnote{Notice 
that $dz$ is not vanishing in $x=\infty$. 
This is consistent with
the fact that the point at infinity is not a branch point in the
Donagi-Witten curve.}.
% is nothing but the point $x=\infty$ of the
%matrix model, and to add $dx$ to $\PP(z)dz$. 
We can construct a differential with the desired properties
by adding $dx$ and $\PP(z)dz$. This easily provides us with
a differential with double poles in $p_i$ with coefficients
$-m(-\tilde n+1,1,...,1)$.
We should also require $dv$ to
have zero periods, since we want to integrate it to give a well
defined meromorphic function $v$ on $\Sigma$.
%The problem of this proposal is
%that the sum of these two pieces has periods, and that would not make a
%good candidate for $dv$, which should not have any periods. 
We can still add a piece with $dz$ without adding any pole. 
This way we end up with a differential $adx-m(\PP(z)+b)dz$, 
where the constants $a$ and $b$ have to be 
determined imposing the vanishing of all periods. 
Only two of the $2N$ conditions on periods are non trivial. 
%A priori this might seem 
%difficult since $\Sigma$ has 2$N$ periods. 
Indeed, the pull-back of 
any $f(z)dz$ has possibly non--vanishing periods only around the two cycles of $T^2$. 
The same statement holds for $dx$; since the periods of $dx$ are all 
equal (\ref{eq:dx}), the only 
independent non--vanishing period is around the cycle $B$, corresponding 
to one of the 
cycle of the base torus. 
Thus, we reduce to only 
two equations, which determine uniquely $a$ and $b$.
The result is
%La parte commentata ha gli $\omega_1$
%%%%%%%%%%%%%%%%%%%%%%%%%%%%%%%%%%%%%%
%\begin{equation}
%  \label{eq:dv}
%  dv = \frac{2 \pi i}{m \omega_1} dx + \frac1{\omega_1^2}
%\Big(\PP(z) -\frac{\pi^2}3 E_2(\tau) \Big) dz\ ,
%\end{equation}
%where $E_2(\tau)$ is a standard Eisenstein series. This can be 
%integrated as
%\begin{equation}
%  \label{eq:v}
%  v= \frac{2 \pi i}{m \omega_1} x + 
%\frac\pi{2\omega_1}\frac{\theta_1'(z|\tau)}{\theta_1(z|\tau)}
%-\frac{\pi^2 E_2(\tau)}{4\omega_1^2}z 
%\end{equation}
%%%%%%%%%%%%%%%%%%%%%%%%%%%%%%%%%%%%%%%
\begin{equation}
  \label{eq:dv}
  dv = 2 \pi dx - m
\Big(\PP(z) +\frac{\pi^2}3 E_2(\tau) \Big) dz\ ,
\end{equation}
where $E_2(\tau)$ is a standard Eisenstein series. This can be 
integrated up to a constant to give \footnote{Here $h_1(z)=\zeta(z)-\frac{\zeta(\omega_1)}{\omega_1} z$ for a torus with periods $2\omega_1$ and $2\omega_2$.
$\zeta(z)$ is the Weierstrass zeta function: it is defined by
having a simple pole in $z=0$ and quasi-periodicity properties $\zeta(z+2\omega_i)=\zeta(z)+2\zeta(\omega_i)$. $h_1(z)$ is periodic along $A$ and
$h_1(z+2\omega_2)=h_1(z)-\frac{\pi i}{\omega_1}$. Other useful
identities are: $\zeta'(z)=-\PP(z)$, $\omega_2\zeta(\omega_1)-\omega_1\zeta(\omega_2)=\pi i/2$, $\zeta(\omega_1)\omega_1=\pi^2E_2(\tau)/12$.}
\begin{equation}
  \label{eq:v}
  v= 2 \pi x + 
m\pi\frac{\theta_1'(\pi z|\tau)}{\theta_1(\pi z|\tau)}
\equiv 2\pi x +mh_1(z)\  .
\end{equation}
This expression is well-defined since $x$ and $h_1(z)$ are
both periodic along the $A$ cycle and the coefficients are chosen
in order to cancel their jump along $B$. 
%This expression is very similar to the one between $k$ and $\tilde k$ in the
%expression of the Donagi-Witten curve as coming from the Lax matrix of the
%elliptic Calogero-Moser system \cite{dhp}. We will comment more about
%this later. 
%Notice that we have not bothered
%to check that the resulting $v$ has the right pole in $p_0$: this is 
%automatically enforced by the vanishing of the sum of residues of the
%differential $vdz$. The implicit assumption in this argument
%is that the number of sheets is really $N$. 
To determine the order $\tilde n$ of the covering, it is now enough to
compute the value of the residue of $x$ at $x=\infty$. 
From equation~(\ref{eq:G}) it follows 
\begin{equation}
\frac{\de}{\de S_i}Gdx=m\frac{dx}{x^2} \ ,
\end{equation}
so that $dz\equiv\Omega= (mN/2\pi)dx/x^2$. It follows that, 
in local coordinates,
$x=- mN/(2\pi \, z)$. This fixes the residue of $v$ near $x=\infty$ to be 
$-(N-1)m$. Thus $\tilde n=N$, and our proof is completed.

Once the algebraic curve is given, eq.~(\ref{eq:v}) determines the
map to the matrix model plane $x$. The function $G$ is also 
uniquely determined by the requirement of having a single pole of order
$N+1$ at $x=\infty$, even though its explicit form can be difficult
to find. Using the results of \cite{dw},  we can determine  the expression 
of the curve and $G$ for small values of $N$. The spectral curve 
(\ref{dwcurve}) can be written as a pair of equations \cite{dw}
\begin{eqnarray}
&&y^2 = (t-e_1)(t-e_2)(t-e_3) \ ,\nonumber\\
&&F_N(v,t,y)=0 \ ,
\label{covering}
\end{eqnarray} 
where the first equation is the standard representation 
of the torus as a cubic, while the second 
is a polynomial of degree $N$ in $v$, giving the $N$-sheeted covering 
of the torus. 
As shown in \cite{dw}, $F$ is also a polynomial in $x$ and $y$.

For example, for $N=2$ and $W'(\Phi)=\Phi^2-A_2$, 
%which is already nontrivial if one considers
%two cuts, 
%by imposing as we said above that $G$ have only a
%single pole of order $N$ in $p_0$, 
we have 
\begin{eqnarray}
&&F_2(v,x,y)= v^2 -t - A_2 \ , \nonumber\\
&&G=y +v^3 -\frac{3A_2}{2} v \ .
\end{eqnarray}
%$A_2$ is the only SU(2) modulus.
Analogously, for $N=3$ and $W'(\Phi)=\Phi^3-A_2\Phi-A_3$
we have 
\begin{eqnarray}
&&F_3(v,x,y)= v^3 -v(3t+A_2) +2y -A_3\ ,\nonumber\\
&&G=v^4 -v^2(6t-\frac{2}{3}A_2) +8vy - 3t^2-\frac{2}{3}A_2 t \ .
\end{eqnarray}
The degree $N$ 
polynomial in $F$ can be related to $W'$ by analyzing the large $x$
behavior of $G$ (see equation \ref{eq:G}).
%{\bf In general}, a way of finding $G$
%would be to get its dependence on $v$ via (\ref{eq:G}) and
%(\ref{eq:v}), and determine afterwards its dependence on $t$ and
%$y$. {\bf Forse ci si pu{\`o} pensare\ldots}
In both cases, $x$ is determined by equation~(\ref{eq:v}). The structure
of these two examples suggests that in general $G$ is a linear combination of
the polynomials $P_k$ defined in \cite{dw}.

For generic $N$ it is probably more convenient to use
an alternative expression for the Donagi-Witten curve \cite{dph},
\begin{equation}
\sum_{n=-\infty}^\infty (-)^n q^{\frac{1}{2}n(n-1)} e^{nz}H(x-nm)=0 \ , 
\label{dh}
\end{equation}
where $H$ is a degree $N$ polynomial, which we can roughly identify
with $W'$. 
Another advantage of this expression is that it is naturally 
written in terms of the matrix model variable $x$. Indeed,
as shown in \cite{dph},
to go from the polynomial equation~(\ref{covering}) to the expression
above, a change of variables $v\rightarrow v-mh_1(z)$ is required, 
which, by equation~(\ref{eq:v}) 
exactly defines the function $x$. 

%In this formalism,
%the function $G$ is explicitly given by
%{\bf formula per G hopefully....} 

\subsection{The general case}
We can also study the general case with arbitrary $n$ and $N$.
In this case, the group is broken to $\prod_{i=1}^n U(N_i)$ and
there are some non-abelian gauge factors at low energy.
For these vacua, we could also introduce, as in \cite{csw2},
integer numbers $b_i$ labeling the type of confinement in each factor.
The numbers $b_i$ appear in the matrix model expression for $W_{eff}$
as \cite{csw2}
\begin{equation}
  \label{eq:Weff2}
  W_{\rm eff}= 
\sum_i \left( N_i \frac{\de {\cal F}}{\de S_i} - 2\pi i\tau S_i\right ) 
- \sum_{i=2}^n 2\pi i b_i S_i\ .
\end{equation}
The minimization procedure then fixes the periods of $\Omega$
to be  $N_i$ and $\tau+b_i$. As before, we conclude that the map
$z:\Sigma\rightarrow \mathbb{C}$ is well defined if we make the
identifications $z\sim z+\tau$ and $z\sim z+t$, where $t$
is the highest common factor of the integers $N_i$ and $b_i$.
It was shown in \cite{csw2} that $t$ defines the index of confinement
of the vacuum.

We see that $\Sigma$ becomes a covering of a torus of modular parameter
$\tau/t$. We can further show that $\Sigma$ is a 
$N/t$-sheeted covering
of the base torus, and express $\Sigma$ as an
algebraic equation. Indeed, the argument  we used to identify the function
$v$ can be repeated almost verbatim in the general case.
The only difference is now the ratio
of the residues in the points $p_i$: 
equation~(\ref{eq:v}) is replaced by
\begin{equation}
v=\frac{2\pi}{t} dx +mh_1(z) \ .
\label{eq:v2}
\end{equation}
Since the behavior of $x$ at infinity is still given by $x= - mN/(2\pi \, z )$,
the residue at infinity is now $N/t$. It follows
that the number of sheets is $\tilde n=N/t$. The genus $n$ 
curve $\Sigma$
is then expressed as an element of the spectral family $F_{\tilde n}(v,z)=0$.
Since the arithmetic genus of a curve of the family $F_{\tilde n}(v,z)=0$ is
$\tilde n>n$, the algebraic curve will be singular and $\Sigma$ will
correspond to its normalization.    

This result deserves some comments.
We know that the relevant geometry 
for arbitrary $n$ and $N$ can be determined as a particular point
in the moduli space of the underlying $U(N)$ $\nn=2$ theory.
In a vacuum with $\prod_{i=1}^n U(N_i)$ only $n$ photons remain massless
and this requires that $N-n$ monopoles are massless and condensate
in order to give mass to all the other degrees of freedom. 
The associated curve has then $N-n$ nodes. The
sub-variety of the moduli space where $N-n$ monopoles are massless
has dimension $n$. The point in this sub-variety associated to the
$\nn=1$ vacuum can be determined by explicitly solving the equations
of motion for the potential $W(\Phi)$. This minimization was
explicitly done in the case of a pure gauge theory in \cite{int}.
In that case,
one can explicitly check that a potential $W(\Phi)$ of degree $n+1$
selects a point in the $\nn=2$ moduli space of the form
\begin{equation}
y^2=P_N(x)^2-1=\prod_{k=1}^{N-n}(x-u_i)^2 (W'(x)^2+f_{n-1}) \ , 
\label{curvepure}
\end{equation}
where the $N-n$ double zeros correspond to a degeneracy of the curve
associated with $N-n$ massless monopoles. The reduced genus $n-1$ curve 
$y^2=(W'(x)^2+f_{n-1})$ is exactly that describing the matrix model geometry.
It would be interesting to repeat this analysis for the case with
a massive hypermultiplet. The $\nn=1$ vacuum $\prod_{i=1}^n U(N_i)$
should be associated with an element of the spectral family
$F_N(v,z)=0$ with $N-n$ nodes, thus degenerating to a genus $n$ surface. 
The matrix model computation suggests that the normalization of such curve 
can be also find by  normalizing a  curve $F_{\tilde n}(v,z)$ of 
Donagi-Witten type
but with a different base torus of modular parameter $\tau/t$.
%It would be interesting to explicitly check this result starting
%from the $\nn=2$ theory.

Finally we would like to comment about the 
cases where some of the cuts coincide.
In such cases, the genus $n$ curve $\Sigma$ degenerates.
The normalization of the resulting curve
has genus equal to the number of cuts. We can for example make contact
with the examples discussed in \cite{dv3,dhks}. These 
correspond to completely massive vacua and are associated with
a single cut. The previous discussion can be repeated with
$n$ replaced by the number of cuts ($n\rightarrow 1$) and
the only non trivial $N_i=N$. We exhibited $\Sigma$ 
as a $1$-sheeted covering (i.e. an isomorphism) of a torus 
of modular parameter $\tau/N$. 
For $n=1$, the Donagi-Witten curve becomes quite
trivial, $F_1(v,x,y)= v-A$, and indeed describes a torus. 
Equation~(\ref{eq:v}) then gives 
$\frac{2 \pi }m  x =
A -\pi  \frac{\theta_1'(z|\tau)}{\theta_1(z|\tau)}
$. 
This is exactly 
the map from the torus to the matrix model plane 
given in \cite{dv3,dhks}. 
%Moreover, equation~(\ref{eq:min})
%adapted to the case of $N$ colors and a single cut gives 
%$\int_B \omega_o =\tau/N$. Therefore, as in \cite{dv3,dhks},
%the matrix model plane is mapped to a torus of modular parameter
%$\tau/N$. 
The field theory interpretation of this result is simple.
Indeed, as it is well known \cite{dw}, the massive vacua
of the $U(N)$ theory with bare coupling $\tau$ 
correspond to points in the moduli space of the $\nn=2$ theory 
where the genus $N$ curve $F_N(v,x,y)=0$ maximally degenerates
and becomes a torus of modular parameter
$\tau/N$.

%Another point deserves a comment. In the general case the mapping $x$ between
%the matrix model plane and the Riemann surface $\Sigma$ 
%has a priori nothing
%to do with the function $G$. In the single cut case and with quadratic $W$
%it so happened that (up to a constant) $G(x)=\PP(z)$, 
%the Weierstrass function on the torus, which is also one of the coordinates 
%$(t,y)$ in the standard embedding. Thus in a way $G$ and $dx$ happened to be 
%related. Though this particular relation cannot be true in general, the 
%properties of $G$ are still enough to find it explicitly after having gone 
%to the $\nn=2$ limit, as we are going to see {\bf in examples}.

\newsection{Resolvents}
\label{sec:res}

The interesting quantities to compute in quantum field theory
are the resolvents \cite{cdsw}
\begin{eqnarray}
R(x)&=&{\rm Tr}(\frac{{\cal W}_\alpha {\cal W}^\alpha}{ x-\Phi})\ ,
\nonumber\\
T(x)&=&{\rm Tr}(\frac{1}{x-\Phi})\ .
\label{QFTresolvent}
\end{eqnarray}
The knowledge of $R(x)$ and $T(x)$ allows to compute all
the vacuum expectation values of operators in the chiral
ring \cite{cdsw}. In this section, we will discuss a possible
identification of $R$ and $T$ with geometrical quantities. 
In the case of pure gauge, $R$ can be related to the meromorphic function
$y$ defining the curve (see equation~(\ref{riemann1})) while $T(x)dx$
becomes a meromorphic differential on the curve \cite{cv,dv3,cdsw,gopa,csw3}.
 
\subsection{Identification of the resolvents} 
We expect that, as in \cite{cdsw}, $R(x)$ is identified with 
the matrix model resolvent 
\begin{equation}
R(x)= S \, \omega (x) \ . 
\label{res}
\end{equation}
In the case of pure gauge, 
this identification descends from  the comparison of the matrix model
loop equations with the Ward identities in field theory \cite{cdsw}.
As shown in \cite{cdsw}, the equations for $R(x)$,  which  
can be deduced in the quantum field
theory using the Konishi anomaly, are formally identical
to the matrix model Ward identities for $\omega(x)$. 
In our case, the Ward identities are more complicated 
to write (some explicit
relations are discussed in the Appendix), but we expect that 
the general philosophy still applies.
%this relation is derived
%by comparison of the matrix model loop equation with
%the quantum field theory Ward identities \cite{cdsw}.

Quantum mechanically, we may expect that
small cuts are opened around the classical eigenvalues of $\Phi$,
analogously to what happens for the matrix model. Integrals of
$R(x)$ around a critical point of $W$ define the condensates 
${\rm Tr}(W_{\alpha}W^{\alpha})_{(i)}$.
The contour integrals of $R$ around a cut 
 are mapped to integrals of
$\omega(x)$ around its cut in the matrix model plane (these cuts are
indicated with dotted lines in figure 2),
which define the quantities $S_i$
\begin{equation}
{\rm Tr}(W_{\alpha}W^{\alpha})_{(i)}=
\frac{1}{2\pi i} \oint R(x) dx =\frac{1}{2\pi i} \oint S \, \omega(x) dx=
S_i\ . 
\label{R}
\end{equation}

%\begin{figure}[h]
%\centerline{{\small\fbox{$x$}}
%\includegraphics[width=13cm,height=4.5cm]{mm2.eps}}
%\caption{\small The $C_i$ contours encircle the cuts of $\omega(x)$,
%denoted by dotted lines.
%The contours $A,A^*,B,B^*$ delimit a simply connected region in the $x$ plane.}
%\label{cuts2}
%\end{figure}
\begin{figure}[h]
\begin{picture}(80,140)(-20,-5)
\put(0,0){{\small\fbox{$x$}}}
\put(15,10){\includegraphics[width=13cm,height=4.5cm]{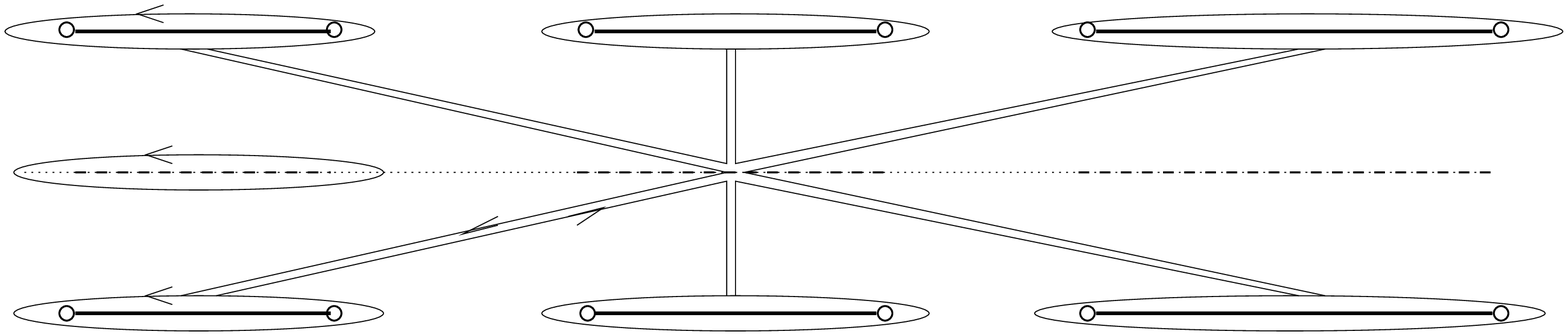}}
\put(40,140){$A$}\put(40,0){$A^*$}\put(40,85){$C$}
\put(115,32){$B$}\put(105,51){$B^*$}
\end{picture}
\caption{\small The $C_i$ contours encircle the cuts of $\omega(x)$,
denoted by dotted lines.
The contours $A,A^*,B,B^*$ delimit a simply connected region in the $x$ plane.}
\label{cuts2}
\end{figure}

From quantum field theory it is also obvious that the integrals of
$T(x)$ around a critical point of $W$ are integers
\begin{equation}
\frac{1}{2\pi i} \oint T(x) dx =N_i \ .
\label{T}
\end{equation}
In analogy with \cite{csw3}, we may expect $T(x)$ to be associated
with a differential on the Riemann surface.
Consider the differential 
\begin{equation}
t(x)dx=\Big[T\Big(x-\frac{i}{2}m\Big)- T\Big(x+\frac{i}{2}m\Big)\Big]dx\ ;
\label{t}
\end{equation}
in a similar way as for the matrix model resolvent, the residue 
cancels from the two $\pm im/2$ pieces, and the differential is 
holomorphic around $x=\infty$. We now conjecture that $t(x)dx$ can be extended
to a holomorphic differential defined on the entire Riemann surface.
Its periods  
around the $A_i$ cycles are $N_i$, by equation~(\ref{T}).
%the usual argument about the jump in the
%resolvent being equal to the distribution ({\bf assumption} about distribution
%of eigenvalues having the same support as the mm one) nothing but the 
%number of eigenvalues of $\Phi$ in the $i$th vacuum, $N_i$. 
These are the same $A_i$ periods as $dz$. 
Since an holomorphic differential is completely specified by its
$A_i$ periods
%it then follows that also the $B_i$ periods
%coincide 
we conclude that
\begin{equation}
  \label{eq:dzFT}
\Omega\equiv dz=\frac{1}{2\pi i}t(x) dx=\frac{1}{2\pi
  i}\Big[T\Big(x-\frac{i}{2}m\Big)- T\Big(x+\frac{i}{2}m\Big)\Big] dx\ .
\end{equation}
It then follows that also the $B_i$ periods
are completely specified (this argument was also 
used in a similar way in \cite{csw3})
\begin{equation}
\frac{1}{2\pi i} \int_{B_i} t(x) dx =\tau \ .
\label{T2}
\end{equation}

The validity of equation~(\ref{eq:dzFT}) is strengthened by the following
formal %but suggestive 
argument. Using the definition of $dz$ and
the identification~(\ref{res}), we have
\begin{equation}
dz=\frac{1}{2\pi}\sum_iN_i\frac{d}{dS_i}Gdx=-\frac{1}{2\pi i}
\sum_iN_i\frac{d}{dS_i}\Big[R\Big(x+i\frac{m}{2}\Big)
-R\Big(x-i\frac{m}{2}\Big)\Big] \ ,
\label{di}
\end{equation}
which, by comparison with~(\ref{eq:dzFT}), leads to the suggestive equation
\begin{equation}
\label{eq:dG}
\frac1N \sum_i N_i \frac\de{\de S_i} {\rm Tr}
\Big( {\cal W}_\alpha {\cal W}^\alpha 
\frac1{x-\Phi}\Big)
={\rm Tr}\Big( \frac1{x-\Phi}\Big) \ .
\end{equation}
%{\bf up to a quantity} periodic under $x\to x+im$. 
%Finally, we also know that the equations for the 
%resolvents $T(x)$ always have the form of a
%derivative of the ones for $R(x)$, due essentially to the ``superfield''
%formalism of \cite{cdsw}. 
%One can compare this procedure with actually
%hitting the equation for the resolvent with ${\cal W}_\alpha{\cal
%  W}^\alpha$ by $\de/\de S$ gives the same equation, and thus
%(\ref{eq:dG}) strongly suggests itself.}
%This equation of course 
%could have formally guessed, and it is remarkable that it gets confirmed by
%geometrical techniques 
%
%In the case of pure gauge, $T(x)$ are then obtained
%by differentiating those for $R(x)$, as a consequence of the
%superfield formalism introduced in \cite{cdsw}.

\subsection{Field theory expectation values} 
We can use $t(x)dx$ to compute the field theory expectation values 
of $\langle{\rm Tr} \Phi^k\rangle$. To this purpose, we can integrate
$x^k t(x)dx$ on a small contour around $x=\infty$,
\begin{equation}
\langle{\rm Tr} \Big[(\Phi+\frac{i}{2}m)^k-(\Phi-\frac{i}{2}m)^k \Big]\rangle=
\int_{\infty} x^k dz \ .
\label{moments}
\end{equation}
We can deform the previous contour integral in the $x$ plane 
until it encircles the cycles $A_i$ and $A_i^*$ (see figure 2), 
obtaining
\begin{equation}
\int_{\infty} x^k dz= \sum_i (\int_{A_i}x^k dz+ \int_{A_i^*}x^k dz)=
\sum_i \int \Big[(x_0^{(i)}+i\frac{m}{2})^k-(x_0^{(i)}-i\frac{m}{2})^k\Big]dz\ ,
\label{moments2}
\end{equation}
where $x_0^{(i)}(z)$ are maps from the $A$ cycle of the base torus
to contours  around the cuts of the resolvents on the real axis
(indicated by dots in figure \ref{cuts2}). 
We call $C_i$ these contours.
Comparing equations~(\ref{moments}) and~(\ref{moments2})
we obtain the useful formula
\begin{equation}
\langle{\rm Tr}\Phi^k\rangle = \sum_i \int_{C_i} x^k dz =\sum_i
\int x_0^{(i)}(z)^k dz\ .
\label{condensate}
\end{equation}
%$x_0(z)$ can be interpreted as the quantum distribution of field theory
%eigenvalues. 
Formula~(\ref{condensate}) was derived in \cite{dhks}
in the case of a single cut.

We can also compute the vacuum value of the effective potential 
in terms of the function $x_0$. This will also strengthen our 
identification of $t(x)dx$ with $dz$.
On shell, we can write
$W_{\rm eff}$ as
\[
\sum_i \left( \int_{A_i}dz \int_{B_i} G\, dx - 
    \int_{B_i}dz \int_{A_i} G\, dx \right)\ .
\]
By Riemann bilinear relations \cite{gh}, this expression is also equal to
Res$_\infty (z(x) G(x)dx)$. The $U$ piece in the definition (\ref{eq:G}) of $G$
is the only one which can contribute to this residue. 
We can at this point
try to invert the proof of the Riemann bilinear relations. To this purpose
we choose a base point, say on the real 
axis of $x$, and modify all the $A_i$ and $B_i$ periods in such a way that
they bound a simply connected region (see figure \ref{cuts2}). 
This way we build a polygon whose sides 
are $A_i$, $B_i$ and their opposites $A_i^*$, $B_i^*$, 
identified in such a way as to reconstruct the Riemann surface $\Sigma$,
similarly to \cite{gh}. In the interior of this polygon 
(a simply connected region) the function $z$ is 
well-defined. The Riemann bilinear relations are then demonstrated by
deforming  a contour integral of $z \, Gdx$ around $x=\infty$ to
the perimeter of the polygon and by exploiting the periodicities of $z$.
If we now substitute $G(x)$ 
with $U(x)$, we obtain extra contributions 
coming from the fact that $U(x)$ is a well defined function 
on the plane $x$ but not on the Riemann surface. But we can now
exploit that to our advantage:
\begin{eqnarray}
\label{eq:Ci}
&&W_{\rm eff}={\rm Res}_\infty (z \,Udx)= 
-\sum_i \left( \int_{A_i+A_i^*} z\,U dx + \int_{B_i+B_i^*} z\,U dx \right)
\nonumber \\
&& = -\sum_i \left( \int_{A_i}(z+\tau)U(x_0+i\frac{m}{2})+\int_{A_i^*} z\,U(x_0-i
\frac{m}{2}) dx - \int_{B_i} U dx \right )\nonumber\\
&& = -\sum_i\left(\int_{C_i}z W'(x)dx - \int_{B_i}U(x)dx\right) \nonumber\\
&& = \sum_i \left(\int_{C_i} W(x)dz\right) + N\int_{-im/2}^{im/2}U(x)dx \ .
\end{eqnarray}
In second line of \ref{eq:Ci}, we first used equation (\ref{eq:U}) and 
the fact that  $U(x)$ is not well-defined on $\Sigma$ 
%and equation (\ref{eq:U}) 
to evaluate the integrals over $A_i$; the integrals over $B_i$
instead just behave as for the usual Riemann argument. 
We have also used that $\int_{A_i}U=0$. We then integrated
by parts the integrals over $C_i$, and then used appropriate integrals of
(\ref{eq:U}) again to put the second piece in the final form.

The final expression for $W_{\rm eff}$ in (\ref{eq:Ci}) is consistent
with the identification~(\ref{condensate}). Indeed, if
$W\equiv\sum_p g_p {\rm Tr}\Phi^p$,
we know $W_{\rm eff}= \sum g_p \langle{\rm Tr}\Phi^p \rangle$. 
So equation~(\ref{eq:Ci}) can be written as
\begin{equation}
  \label{eq:up}
  \sum_i \int_{C_i} W(x)dz 
\sim \langle\sum_k g_k{\rm Tr} \Phi^k\rangle\ ,
\end{equation}
modulo pieces which, for a given $W$, 
depend on $g_k$, but not on $\tau$ and on the
choice of a particular vacuum. Formula
(\ref{eq:Ci})
was derived in a different way in the one-cut case in \cite{dhks}.

Formula~(\ref{condensate}) gives a prescription
for computing the quantities $<{\rm Tr}\Phi^k>$ purely in terms
of matrix model data; the function
$x_0(z)$ can be interpreted as the quantum distribution of field theory
eigenvalues. We should note, however, that there is
an ambiguity in the definition of $<{\rm Tr}\Phi^k>$
in field theory. The condensates can be computed as the order
parameters $u_k=<{\rm Tr}\Phi^k>_{\nn=2}$
of the $\nn=2$ vacuum that is selected by the
potential $W(\Phi)$. In presence of a mass $m$, $u_k$ can mix
with all the other order parameters $u_j, j<k$ \cite{dw,ad}.
The results for the condensates obtained with different methods
could be related by a change of basis in the $u_i$; consistency
requires the coefficients of such redefinition to be
vacuum independent.
In the case of a single cut, it was explicitly checked in \cite{dhks} 
that the condensates computed with the matrix model prescription 
are indeed related by a vacuum-independent redefinition to
the condensate computed using the Donagi-Witten curve.

\vskip .2in \noindent \textbf{Acknowledgments}\vskip .1in \noindent   
We would like to thank Annamaria Sinkovics and 
Stefan Theisen for interesting discussions.
This work is partially supported by the EU contract
HPRN--CT--2000--00122. M.P. is supported by the European Commission Marie
Curie Postdoctoral Fellowship under contract number HPRN--CT--2001--01277.
A. Z. are partially supported by INFN and MURST, and
by the European Commission TMR program HPRN-CT-2000-00131,
wherein he is associated to the University of Padova.

\section{Appendix: Loop equations}

The identifications of Section 4 could be strengthened
by comparing the matrix model loop equations with the Konishi
anomaly equations in field theory. A complete set of loop equations
uniquely determining the resolvents would also give a convenient
description of the off-shell theory. In the case of pure gauge,
the loop equations for the matrix model give relation~(\ref{riemann1})
for $y=W'(x)-2\omega(x)$, which uniquely determines the
Riemann surface associated with the matrix model; this result
is also valid off-shell. In the case of $\nn=1^*$, the
loop equations are more complicated. 
Here we will make a first step in the study of the
loop equations and we
derived a suggestive relation satisfied by the resolvent.
In the following, we will refer to the Ward identity for the matrix model, 
which are identities for $\omega(x)$. As shown in \cite{cdsw},
identical equation for $R(x)$ can be deduced from the quantum field
theory Ward identities. Identities for $T(x)$ are then obtained
by differentiating those for $R(x)$, as a consequence of the
superfield formalism introduced in \cite{cdsw}.

%We try now to extract more informations about the matrix model resolvent,
%of which (\ref{eq:rs}) was a priori not all, as commented there. 
The best
way to find the loop equations is to consider the matrix model {\sl before}
integrating out $\hat X$ and $\hat Y$ (\ref{eq:supot}), 
and making appropriate changes of coordinates \cite{schnitzer,ragazzi}.
The ones we consider here are
\[
\delta\hat \Phi= \frac1{x-\hat\Phi}\ , \qquad 
\delta \hat X= \frac1{x-\hat\Phi+i\frac m2}\,\hat X\,\frac1{x-\hat\Phi-i\frac
  m2} \ .
\]
The equations that we get from these are
\
\begin{equation}
\label{eq:lxy}
\begin{array}{ccl}\vspace{.2cm}
&& \omega^2(x)= - {\rm Tr}\Big[\Big(W'(\hat \Phi)+[\hat X,\hat Y]\Big)\frac1{x-\hat\Phi}\Big] \ ,
\\ 
&& \omega\Big(x+i\frac m2\Big)\omega\Big(x-i\frac m2\Big)
={\rm Tr}\left(\hat Y\hat X\frac1{x-\hat\Phi-i\frac m2}-\hat X\hat Y\frac1{x-\hat\Phi+i\frac m2}\right)\ . 
\end{array}
\end{equation}

By considering repeated translations of (\ref{eq:lxy}) by $\pm im/2$, 
we can find a combination in which $\hat X$ and $\hat Y$ disappear:
\begin{equation}
\label{eq:loop}
\sum_{n=-\infty}^{\infty} \Big[\omega\Big(x+i(n+2)\frac m2\Big)-
  \omega\Big(x+in \frac m2\Big)\Big]^2=
2\sum_{n=-\infty}^{\infty}{\rm Tr} \Big[\frac{W'(\hat \Phi)}{x-\hat
  \Phi+in\frac m2} \Big]\ .
\end{equation}
One can actually write this equation in terms of $G$ by completing a
square. 
Defining  $X_n(x)\equiv X(x+in\frac m2)$ for any function $X(x)$, we have
\begin{equation}
  \label{eq:loopG}
  \sum_{n=-\infty}^{\infty} G_n^2=\sum_{n=-\infty}^{\infty} U_n^2 - 
2\sum_{n=-\infty}^{\infty} f_n \ ,
\end{equation}
where we have defined 
$f(x)= {\rm Tr} [W'(\hat \Phi)-W'(x))/(x-\hat \Phi)]$. 
The formal analogy
with the pure gauge case is evident. In that case, the loop equations are
$y^2=W'^2+f$. In our case, $G$ plays the role of $y$ and $U$ the role
of $W'$. This would become of course more
than an analogy if one considers the limit $m\to \infty$.

It is not clear whether equation (\ref{eq:loopG}), involving
and infinite series of shifts, is well-defined. If so, it
could be interpreted as an equation determining $G$, and implicitly 
$\Sigma$, in dependence of the polynomials $U$ and $f$. It would
be interesting to investigate whether $G$ is uniquely determined 
by such equation. Similar issues were studied in \cite{ber}.
It would be also interesting to study 
equation~(\ref{eq:loopG}) after minimization, where it should be
related to the Donagi-Witten curve.
The form of equation~(\ref{eq:loopG}) is very suggestive in comparison to the
Lax matrix form of the Donagi-Witten curve given in equation~(\ref{dh}).
%situation, as already noticed above, after (\ref{eq:v}). The formal
%similarities of (\ref{eq:loopG}) with the equation for the curve in
%the $m\to\infty$ and $\nn=2$ limit lead one to conjecture that this is
%the equation for the $\nn=1$ curve.
We leave the investigation of all these issues to future work. 

%The geometry of the
%off-shell theory, which is a $n$-parameter deformation of the
%Donagi-Witten curve, is certainly interesting
%in itself and deserve further investigation. 

\end{document}